# NIR Schottky Photodetectors Based on Individual Single-Crystalline GeSe Nanosheet

Bablu Mukherjee, Yongqing Cai, Hui Ru Tan, Yuan Ping Feng, Eng Soon Tok*, and Chorng Haur Sow*

Department of Physics, 2 Science Drive 3, National University of Singapore (NUS), Singapore-117542

## Abstract

We have synthesized high-quality micrometer-sized single-crystal GeSe nanosheets using vapor transport and deposition techniques. Photoresponse is investigated based on mechanically exfoliated GeSe nanosheet combined with Au contacts under a global laser irradiation scheme. The nonlinearity, asymmetric, and unsaturated characteristics of the I-V curves reveal that two uneven back-to-back Schottky contacts are formed. First-principles calculations indicate that the occurrence of defects create in-gap defective states, which are responsible for the slow decay of the current in the OFF state and for the weak light intensity dependence of photocurrent. The Schottky photodetector exhibits a marked photoresponse to NIR light illumination (maximum photoconductive gain ~ $5.3 \times 10^2$ % at 4V) at a wavelength of 808 nm . The significant photoresponse and good responsivity (~3.5 A $W^{-1}$) suggests its potential applications as photodetectors.

# Introduction

Semiconducting materials in low dimension play an important role in modern materials and science.[1,2] Zero dimensional (0D) and one dimensional (1D) semiconducting nanostructures have been extensively studied with the aims to develop novel application using these materials.[3,4] Among these studies, two dimensional (2D) nanostructures have not been extensively explored. Recently, metal chalcogenides nanostructures with layered structures of group IV elements (GeS, GeSe, $GeSe_2$, SnS, SnSe, etc.) have attracted strong interest.[5-8] The Ge based layered IV-VI nanostructures (i.e., GeSe, GeS) are potential alternatives to the lead chalcogenides due to the advantages of their relatively higher stability and environmental sustainability.[9] Ge based chalcogenide materials are made up with earth abundant elements and they are less toxic. Most recently Li *et al.* reported the vapour deposition growth of GeS chalcogenides nanosheet, where they have focused their study on the fundamental understanding of the synthesis of 2D chalcogenides nanosheet.[5] Semiconducting chalcogenides allowing exfoliation into atomically thin layers have received increasing attentions for showing properties that are complementary to graphene due to the existence of finite band gap. The interesting property of strongly layer-dependent band gap paves an avenue for fabricating high efficient electronic and optelectronic devices, such as multijunction solar cells[10] and layer-dependent photodetector.[11] Recent study has observed photothermoelectric effect in single-layer $MoS_2$ and an orders-of-magnitude larger Seebeck coefficient than that of graphene.[12] For fabrication of photo-based and electronic devices, it has been argued that multilayer nanosheets are more preferred than the monolayer due to multiple conducting channels, a wider spectral response, and larger carriers density compared to the monolayer material.[13] Thus synthesizing new 2D materials with higher mobility and photoresponsibility with massively-sized crystalline nanosheets is highly desired. Materials in 2D geometries can avoid the limitations for 1D nanostructures. Thus 2D nanomaterials are found to be compatible with established device designs and processing approaches in semiconductor industries.

GeSe is a narrow band gap (1.08 eV) IV-VI p-type semiconductor with layered crystal structure.[14] It has a high anisotropic crystallization in layered structure with layers in parallel to growth direction. Low dimensional single-crystalline germanium monoselenide (GeSe) comb structures and their p-type semiconducting transport properties as well as photo-switching behaviour have been studied.[15] Colloidal synthesis of GeSe nanobelts has been reported with electrical measurements of single nanobelt.[16] The previous studies give an overview about the electrical conductivity measurements parallel and perpendicular to the layer planes of p-type bulk GeSe single crystals and the photoconductivity spectral response of bulk GeSe single crystal.[17,18] Current synthesis strategies for GeSe nanosheets rely on solution based process, which may have a solvent residues adsorbed at the surface that are difficult to remove. To the best of our knowledge, there are no reports of such investigation on the optoelectronic properties of 2D GeSe nanosheet based devices with NIR wavelength excitation. In this paper, we report the synthesis of high quality single-crystalline micron sized 2D GeSe nanosheets as well as bulk flakes using vapour deposition techniques. GeSe nanostructures with different morphologies were obtained during synthesis. This approach gives rise to very clean nanosheets with interesting optoelectronic properties and thus rendering them potentially useful candidate for device application.

As GeSe material shows graphite-like layered structures with strong intralayer covalent bonding and weak interlayer van der Waals forces, we use mechanical exfoliation technique to put few layers GeSe nanosheets on $SiO_2$/Si substrate. Patterned electrodes of 200 nm Au were fabricated via standard photo-lithography and sputtering of metals to obtain 2D nanosheet based GeSe photodetector devices. This paper reports for the first time, a large photo-response from single crystalline 2D GeSe nanosheet under NIR laser irradiation. We have performed systematic studies of photoconductivities of the two-point probed device under laser light illumination with photon energy that is above the bandgap of the material. These results provide an insight into the electrical and photoresponse properties from 2D GeSe nanosheet based photodetector. It is found that Au electrodes formed uneven Schottky barriers (SBs) with GeSe nanosheet. The SBs formation was attributed as the main contributing factor to the photoresponse of this metal-semiconductor structure. Weak light intensity dependence of photocurrent and the slow decay of the current in the OFF state were observed in the Schottky photodetector, which could be due to presence of the deep trap states associated with various possible defects in the system. First principles calculations based on density functional theory (DFT) indicate the presence of shallow and deep defective states in the band gap. Thus we have correlated the experimental results with the theoretical calculations. Overall, we have demonstrated the efficiency of GeSe nanosheet-based Schottky photodetector, which shows good response in NIR light illumination (wavelength ~ 808 nm).

## Experimental Section

GeSe of few-layer thickness was fabricated using a two-step process, involving chemical vapor deposition (CVD) based synthesis of GeSe bulk flakes and mechanical exfoliation of the bulk flakes onto SiO$_2$(300 nm)/n-Si substrates using scotch tape based technique. We prepared few-layer GeSe samples from bulk GeSe using a scotch-tape mechanical exfoliation technique similar to that employed for the production of graphene.[19] The exfoliated GeSe nanosheets were characterized by Raman spectroscopy and atomic force microscopy. Single GeSe nanosheet based devices were prepared using standard photolithography techniques, where Au (200 nm) film deposition and lift off were performed to pattern electrodes on the individual GeSe nanosheet. The current-voltage (I-V) and all electrical characteristics of the GeSe nanosheet device were measured using Keithley 6430 source-measure unit. The photoresponse of individual nanosheet photodetectors was measured using continuous wave laser beam from the diode laser (808 nm, EOIN) directly irradiated onto the NB device through a transparent glass window of the vacuum chamber (inside pressure $10^{-3}$ mbar) and the device was electrically connected through a pair of vacuum compatible leads to the source meter.[20] In Global illumination, the laser source is directly mounted on the top of the device such that the different parts (i.e. electrodes, electrodes-nanosheet junctions/interface, and top surface of the nanosheet) of the device are uniformly illuminated with the laser source.[21]

The nanosheets were synthesized in a horizontal single-zone tube furnace using a chemical vapor transport and deposition technique with the following two-step process. Step 1: the mixture of pure Ge powder (Sigma Aldrich, purity 99.99%), Se powder (Sigma Aldrich, purity 99.99%) in molar ratio 1:1 was grounded in a crucible for 30 min. The powder was then placed in a boat, which was inserted inside the vacuum sealed furnace. The furnace was evacuated to a base pressure of 10 mTorr and then flowed with pure Ar gas (flow rate: 200 standard cubic centimeters per minute, SCCM), which was controlled by a mass flow meter. After the pressure has stabilized, the furnace was heated to 480 $^o$C at a rate of 1 $^o$C/min and maintained for 4 hr. The furnace was naturally cooled to room temperature upon the reaction being terminated and the sintered powder was collected. Step 2: 20 mg of this sintered powder was used as a source material for GeSe nanosheet synthesis. 3 cm × 2 cm cleaned Si substrate and a small alumina boat containing the sintered powder were loaded into an one-end open quartz-glass tube with a separation of 28 cm and the boat was placed near to the close end of the tube. The tube was inserted in the vacuum sealed horizontal furnace such that the source material was positioned at the center of the heating zone and the substrate was located down-stream of the source powder. Then the furnace was evacuated to the base pressure of 10 mTorr and subsequently was filled with pure Ar gas (flow rate: 100 SCCM). After the pressure of Ar gas has stabilized to 2 Torr, the furnaced was heated to the targeted temperature of 640 $^{\circ}$C (heating rate: 30 $^{\circ}$C/min) and maintained for 30 min. After the deposition, the furnace was naturally cooled to room temperature and the substrate was collected.

The morphology, structure and chemical composition of the as-synthesized nanostructures were characterized using field emission scanning electron microscopy (FESEM, JEOL JSM-6700F), transmission electron microscopy (TEM, JEOL, JEM-2010F, 200 kV), energy-dispersive X-ray spectrometers (EDX) equipped in the TEM, X-ray diffraction (X'PERT MPD, Cu K$\alpha$ (1.542 Å) radiation and the Raman spectrum was investigated with a Renishaw system 2000 Raman spectrometer under excitation by 514.5 nm Ar$^+$ laser.

X-Ray photoelectron spectroscopy (XPS, Omicron EA125 analyzer) measurements were carried out using a monochromatic Mg K$\alpha$ source (1253.6 eV) to determine the composition of the as synthesized product. Absorption spectrum was obtained using a UV-Vis spectrophotometer (Shimadzu UV-3600). The Veeco D3000 NS49 AFM system is used to measure thickness of the nanosheets. In order to determine incident photon to converted electron (IPCE) ratio, a monochromator (iHR320, HORIBA), which is capable of producing light with wavelengths ranging from 400 nm to 1600 nm and the light intensity was calibrated with a high sensitivity and spectrally flat pyroelectric detector (Newport, model 70362).

# Results and Discussion

The synthesis of high quality, crystalline semiconducting chalcogenides nanomaterials is desired in nanostructured materials research. Chalcogenides nanostructures can be produced in a wide diverse morphology. The synthesis of GeSe structures was carried out using vapor transport and deposition techniques as described in the experimental section.

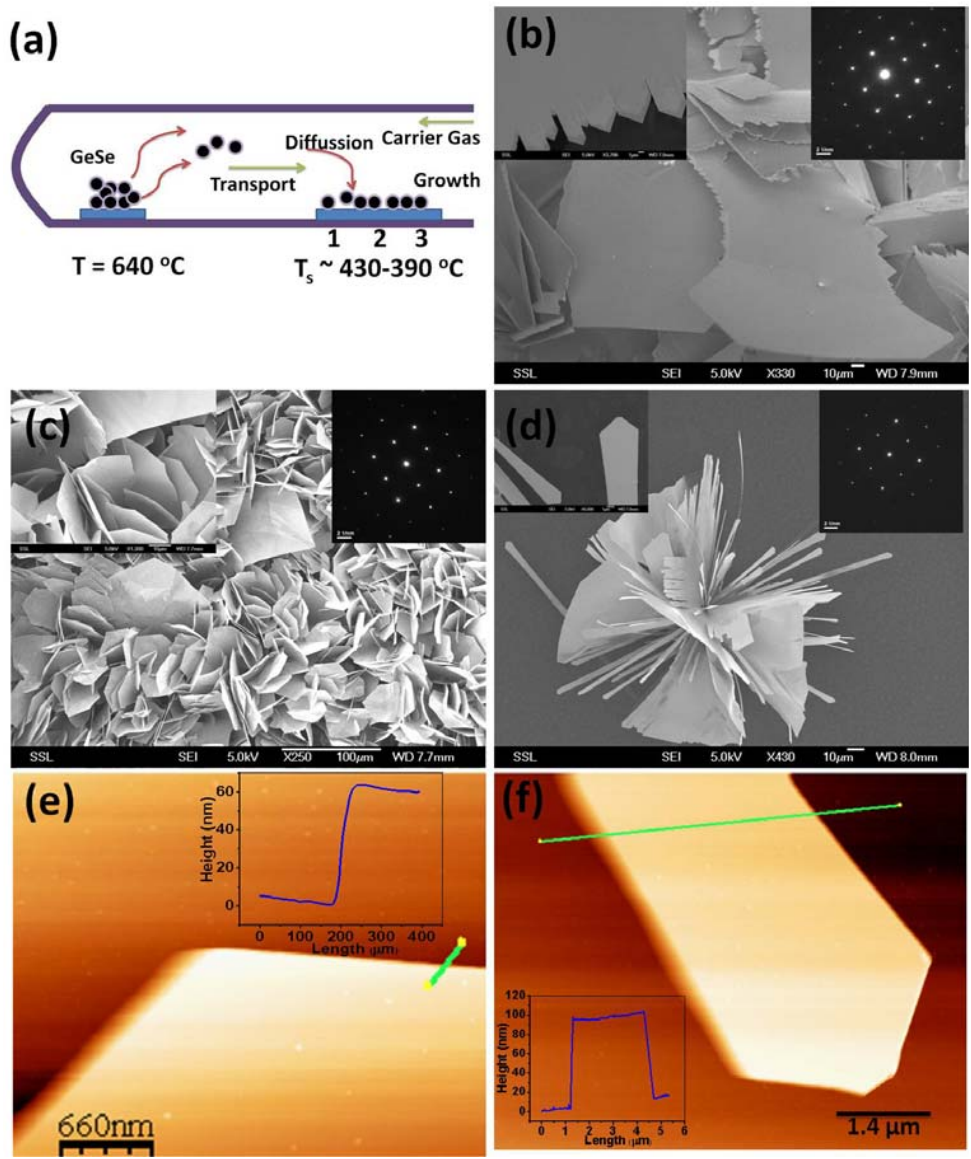

Figure 1. (a) Schematic view of dynamics behavior during the synthetic process. (b,c and d) Low-magnification SEM images of as-grown GeSe nanosheets on the areas indicated by 1, 2 and 3 in (a), respectively, for 30 min growth process on a Si (100) substrate. The insets at the right corners of (b, c and d) are SAED patterns for representative nanosheets. The insets at the left

corners of (b, c and d) are the magnified SEM images of GeSe nanosheets. (e and f) AFM image of GeSe nanosheet and the height profile corresponding to the solid line.

The synthesis and growth mechanisms of nanostructures including nanowires, nanobelts, nanotubes, using vapor deposition techniques have been extensively discussed.[22] However those growth mechanisms are not applicable for vapor deposition growth of crystalline 2D nanosheets. We have not used any catalyst to synthesize the structures and no droplets were found on the as grown flat 2D nanosheets. This suggests the growth was free from any catalyst and nor there any self-catalyst component. Previously GeSe comb structures have been reported, where the authors attributed the formation of GeSe comb structures via an atmospheric vaporization-condensation-recrystallization (VCR) and the formation of wire-fingers from a body plate to the naturally generated structural defects (i.e. screw dislocations, valley-shaped defects, etc.) during synthesis.[15] The nanosheets grew in the furnace directly on Si substrate placed at down-stream position of the source material with lower temperature. The dynamic behavior of synthetic process could be as follows. At first the sublimation of the source powder into gaseous GeSe molecules and transport through the Ar carrier gas flow as illustrated in the schematic Figure 1a. This is followed by condensation of the gas molecules onto the receiving Si substrates placed at the temperature ranges 390 $^{o}$C to 430 $^{o}$C. Constant flux of source molecules adsorbed to growth sites and recrystallization process help to grow the nanosheets. Figure 1(b,c and d) shows scanning electron microscope (SEM) images of typical as-grown micron sized GeSe nanosheets on Si(100) substrate on the areas indicated (in Figure 1a) by 1, 2 and 3, respectively. The different nanostructures with sheet-like or, belt-like morphologies are observed on the substrate. The density and the size of the nanosheet are high at the upstream edge (region 1) of the Si substrate. The upstream edge (region 1) of the substrate facilitates the larger supersaturation of GeSe vapor to condensate into micron sized nanosheet structures. VCR process and vapor-mediated self-assembly process, involve the formation of reactant gaseous materials at high temperature followed by the condensation of the vapors on the substrate at low temperature region, from which precursor-specific recrystallization into various low-dimensional structures are induced. We found smaller size GeSe nanosheet with lower density (Figure 1d) grows at lower temperature position as marked by region 3 of Figure 1a. The concentration gradient of GeSe vapor (the diffusion flux) decreases at lower temperature region (region 3, large distance from the source) could be the contributing reason for the observation of such different crystalline nanostructures. Lower deposition temperature may induce condensation of lesser GeSe vapor and then recrystallization into small vapor flakes. The nanosheet is highly anisotropic in morphology and the lateral dimension can reach over 4-160 μm (Figure 1d), while the thickness is in the range 60-140 nm (Figures 1 e,f), giving rise to a size/thickness ratio of ~ 100-1000. Atomic force microscope (AFM) characterizations indicate that the surface of the nanosheet is reasonably smooth. We can find steps at a height of 2-5 nm at the surface of the nanosheet.

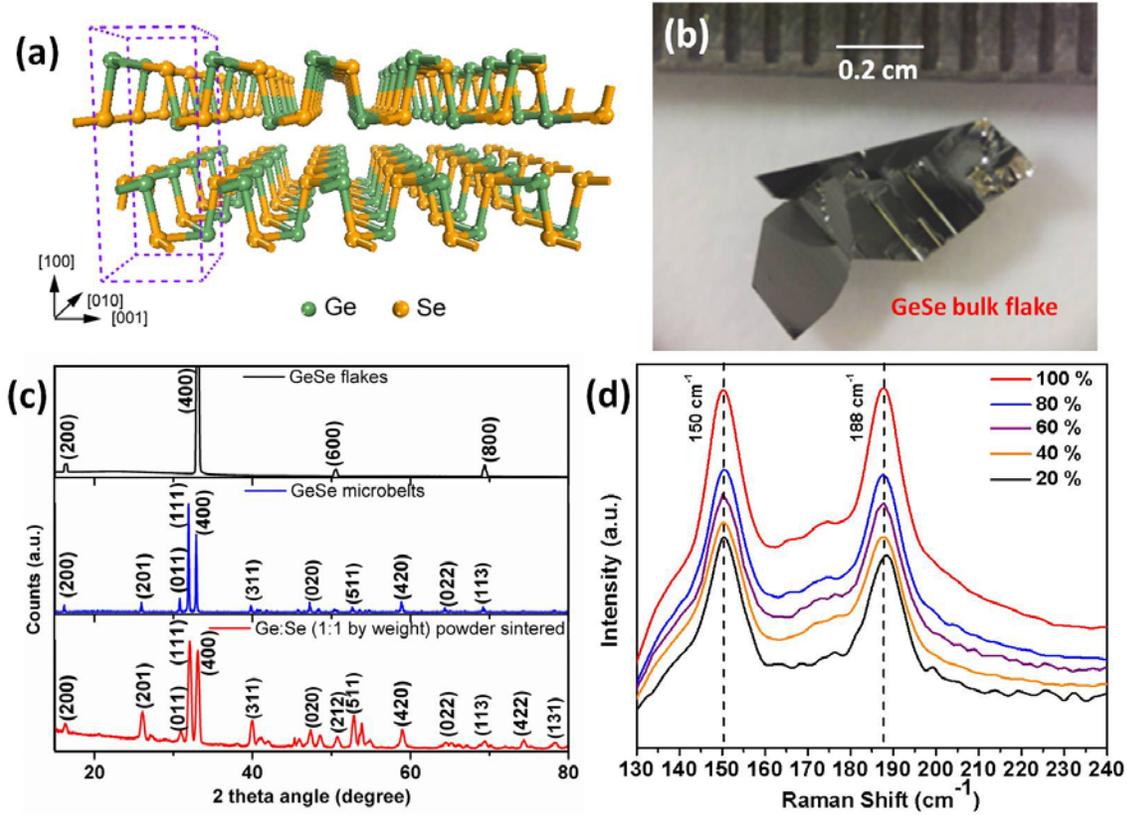

Figure 2. (a) Atomic structural model of the GeSe double layers for the (100) surface. (b) Optical micrograph of thick GeSe bulk flake. (c) XRD patterns of GeSe precursor powder (red line), micron sized nanosheets (blue line) and single bulk flake (black line). (d) Raman spectra for GeSe nanosheet with the intensities of laser excitation. The probe excitation light (λ~ 514.5 nm, 50× objective lens, laser spot size on the film ~ 3 μm) was exposed about 10 s.

Figure 2a represents three dimensional view of the structure of GeSe, which indicates the crystals of GeSe are composed of vertically stacked layers hold together by weak van der waals force. Figure 2b shows the optical micrograph of thick GeSe bulk flake, which was synthesized for long growth duration (growth duration 3hr) under same growth conditions as described in experimental details. These bulk flakes were grown for longer growth duration, which was used to obtain nanosheets on $SiO_2$/Si substrate. Subsequently the bulk flakes were used in mechanical exfoliation method to fabricate single nanosheet based photodetector devices. The detailed crystal structure of the sintered source powder, GeSe nanosheets and GeSe bulk flake was characterized by X-ray diffraction (XRD) patterns (Figure 2c). The GeSe nanosheets and bulk flakes turned out to have an orthorhombic unit cell with a Pnma (62) space group (JCPDS No. 48-1226, a: 10.84 Å, b: 3.834 Å, c: 4.39 Å, β: 90, α phase), which is identical to that of the sintered GeSe source powder. These results indicate that the sintered source powder was not pyrolyzed during sublimation into gaseous GeSe molecules formation and the unit molecule composing the GeSe nanosheet retained the intrinsic source molecular structure. Figure 2d is the Raman spectrum of GeSe nanosheet with different excitation laser power. The peaks at 150 and 188 cm$^{-1}$ can be assigned to the transverse optical (TO) mode in the $B_{2u}$ symmetry and in the $A_g$ symmetry of GeSe, respectively.[23,24]

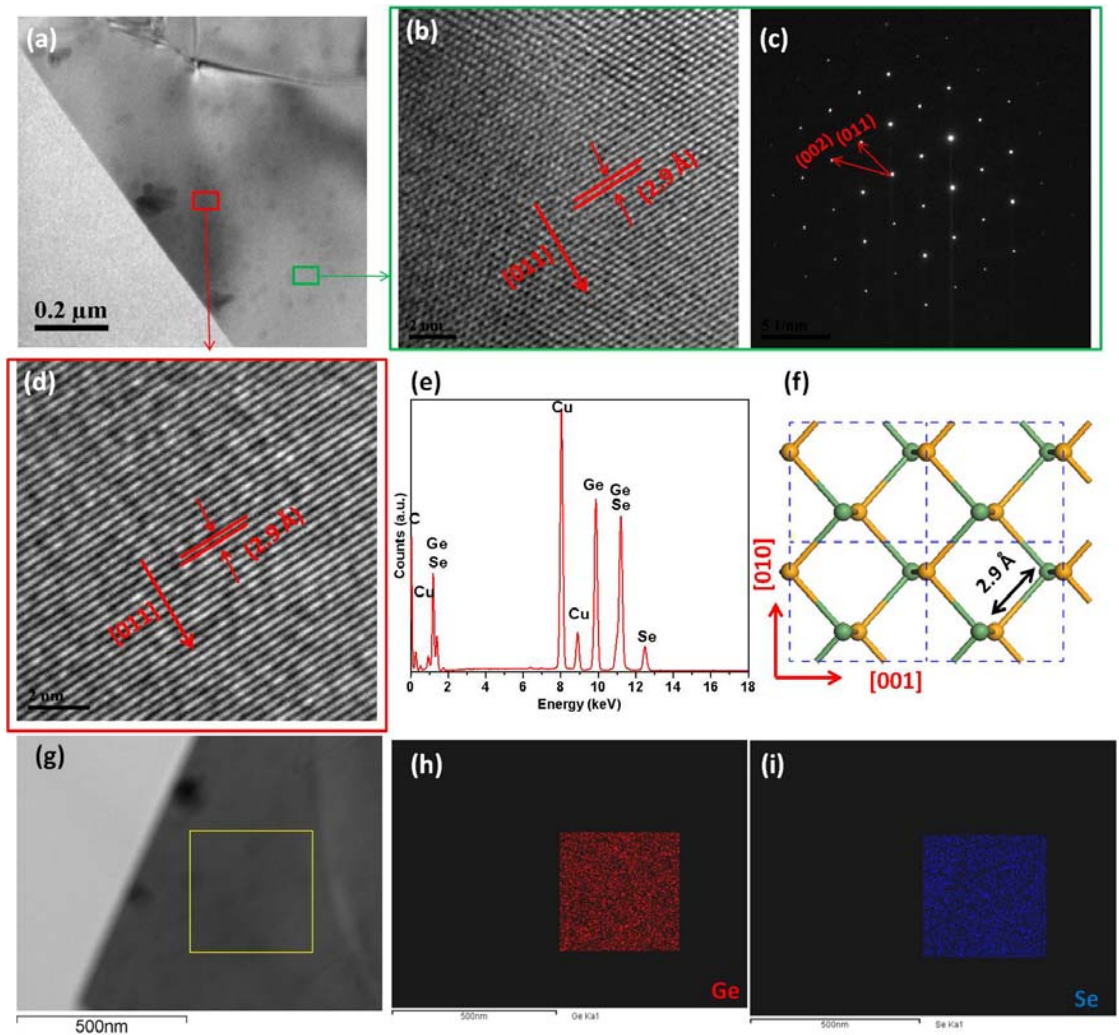

Figure 3. (a) Representative TEM image of a micron sized GeSe nanosheet . (b-c) HRTEM image and SAED pattern obtained from the region that is highlighted by green box. (d) Higher magnification TEM image of the region that is highlighted by the red box, revealing that single crystals is achieved even at a length up to a micrometer. (e) EDS spectrum. (f) Crystallographic view of GeSe molecules on the (100) plane indicating the growth direction of [011] matching the lattice image in (b) and (d). Pale yellow balls and aqua balls represent Ge and Se, respectively. (h, i) EDS map of the region (highlighted by the yellow box in Figure 3g) displaying the uniformly distributed elements of Ge (h), Se (i).

Transmission electron microscopy was carried out to provide better insights into the crystalline order of the micron sized GeSe nanosheet. Selected-area electron diffraction patterns (SAED) and lattice images were taken using a high-resolution transmission electron microscope (HRTEM) at different portion of mechanically exfoliated nanosheet as marked by red and green box in Figure 3a. As shown in Figure 3(b, d) the lattice fringes are exactly identical, which implies that the same crystal structure all over the exfoliated nanosheet. The chemical composition of as-prepared nanosheet structured GeSe was determined to be Germanium and Selenide with an atomic ratio of 1:1 by using EDX (Figure 3e). The carbon and copper come from the thin carbon film and copper mesh of TEM sample, respectively. The SAED (Figure 3c) and HRTEM results indicate that the nanosheet is single crystalline. The basal plane of the nanosheet is always (100), as indicated by the SAED and HRTEM images. It is also observed from Figure 3f that the GeSe molecules pairs are connected on the (100) plane. The elemental maps of the highlighted box in Figure 3g are displayed in Figure 3(h,i) shed the light on the distribution of each constituting element and further verify the entire uniformity of the elements distribution throughout the structure. Furthermore, XPS spectra (Figure S1) confirm the quality, purity and absence of any unwanted chemical residue on surface of the nanostructures.

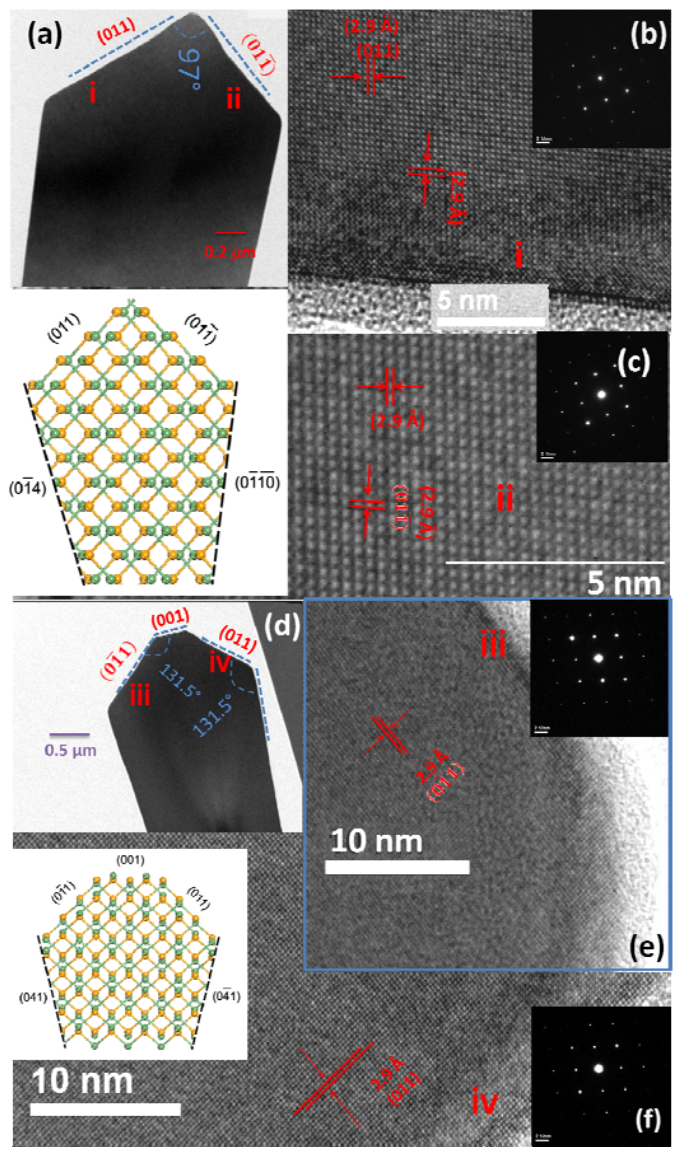

Figure 4(a,d) Low magnification TEM image of single GeSe nanosheet with facets indexed. (b, c, e and f) HRTEM image of the single GeSe nanosheet taken from different parts of the nanosheet as marked by i, ii, iii, and iv in (a, d). Insets in HRTEM images show corresponding SAED pattern. Structural models of representative sheets are shown in the insets in (c) and (f).

Figure 4(a,d) shows the low magnification TEM image of as-grown GeSe single crystal, which exhibits nanosheet-like morphology. The nanosheets appear to be truncated rectangular and triangular shaped. By analyzing the HRTEM images and related SAED patterns, we have identified and indexed the facets of the nanosheets. In addition, we have built structural models for two different nanosheets as shown in the inset in Figure 4(c,f). High-index crystalline planes could be found at the exposed facets, such as, $(0\bar{1}4)$ and $(0\bar{1}\bar{1}0)$ for triangular shapes (Figure 4a)

nanosheet, (041) and (0$\bar{4}$1) planes for truncated rectangular shapes (Figure 4d). The SAED patterns (inset of HRTEM images of Figure 4(b,c,e and f) taken from a single micrometer-sized nanosheet show a spot pattern that is consistent with a single crystal nanosheet having the surface normal oriented along the [100] direction, which is also in agreement with the [100] sheet orientation observed by XRD in Figure 2c. This is dictated by the nature of GeSe material, which is layered material and (100) is the most energetically stable plane. For a better insight into the facets of the single GeSe nanosheet, high resolution TEM images were taken at four different positions as marked by i, ii, iii and iv in Figure 4(a,d). The HRTEM images at positions i and ii of Figure 4(a) exhibit lattice spacings of 2.9 Å. The measured intersection angle between the side facets is approximately 97° for triangular shaped GeSe nanosheets. This indicates the triangular GeSe nanosheets are consistent with the {011} set of planes of the orthorhombic crystal structure and the nanosheets are bound by {011} facets. The measured interfacial angles between the facets are indicated in Figure 4d, which is matched with the calculated interfacial angles according to single-crystal structure. HRTEM image (Figure 4e) shows the lattice spacings of 2.9 Å at position (iii) and the measured angle between the side facet and top facet is approximately 131.5°, which indicates the micrometer-sized nanosheet has {0$\bar{1}$1} and {001} facet at position (iii) and top surface, respectively. The HRTEM image (Figure 4f) at position (iv) shows the lattice spacings of 2.9 Å, which is consistent with the {011} set of planes of the orthorhombic GeSe crystal structure. This refers that the micrometer-sized nanosheet has {011} facet at position (iv). The HRTEM images at different positions and the SAED pattern indicate that the grown crystal has (100) plane parallel to the surface of the nanosheet.

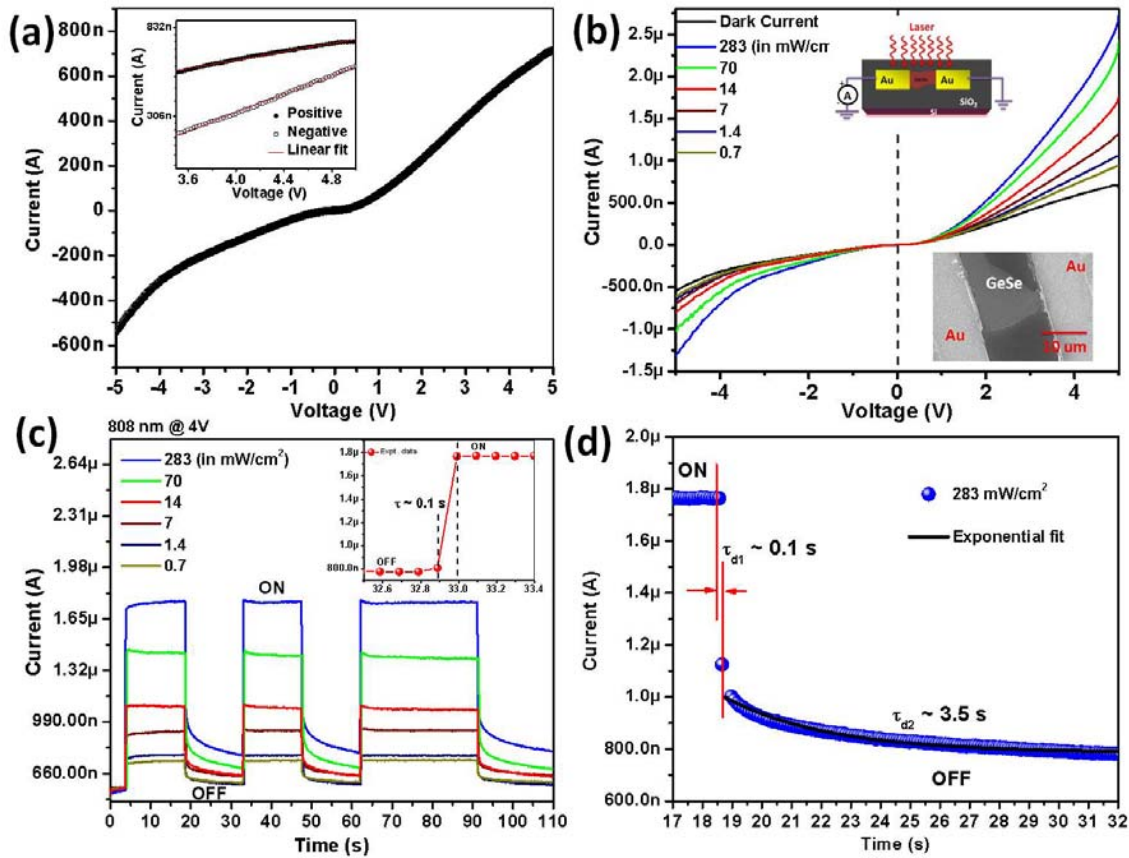

Figure 5(a) Typical I-V curve of Au/GeSe nanoflake/Au in the voltage range (-5 to +5 V) under dark condition. Inset at top-left shows fitted ln(I) vs V curve under dark condition. (b) The performance of GeSe nanosheet-based photodetector device under 808 nm-light illuminations. The inset (top-right) shows the schematic presentation of global irradiation of laser light onto the device during photocurrent measurements. The inset (bottom-right) represents the SEM image of the fabricated GeSe nanosheet-based device. (c) Time dependent photocurrent response of the device to laser light illuminations with different light intensities under vacuum ($4\times10^{-3}$ mbar) at fixed bias of 4V. Inset shows the enlarged views of a 32.6-33.4 s range (from light-off to light-on transition) showing response time ~ 0.1 s. (d) Time-response curve analysis: the decay curve when the GeSe nanosheet device was illuminated at fixed 4V bias. Solid lines represented the fitted curves with the decay equation (1).

Semiconducting nanomaterials often show higher conductivity when irradiated by photons with energy higher than the band gap of the materials.[25] We have tested nanosheet Schottky devices with broad laser beam with photon energy above band gap (laser wavelength: 808 nm, energy ~ 1.5 eV) of the nanosheet. Figure S2 shows the photoresponse of bulk crystalline GeSe under 808 nm light irradiation. The photoresponse of the GeSe bulk flake device in wide range of wavelength scan is shown in Figure S3(a). The reflection spectrum of the micron sized GeSe nanosheets sample is shown in Figure S3(b), which indicates a good absorption of the sample in 808-nm NIR wavelength. Typical I-V curve of Au/GeSe nanoflake/Au in the voltage range (-5 to +5 V) under dark condition is shown in Figure 5a. The dark I-V curve shows asymetric nonlinear characteristics. Due to different nanosheet width at the contacts, the nanosheet forms different contact areas with Au electrodes, which gives uneven contacts and we believe this leads to asymmetric IV response. The nanosheet makes two uneven Schottky barriers (SBs) due to different couplings at the two contacts with the Au electrodes and thus resulted in the asymmetrical I-V characteristics.[26,27] In this metal-semiconductor-metal (MSM) device structure, the current is dominated at positive bias voltage as seen from Figure 5a. From the inset at top-left of Figure 5a, it can be seen that ln (I) is linear with V in the selected high bias range for both positive and negative currents, which indicates the I-V characteristics of back-to-back SB structure. The measured I-V characteristics were obtained under dark condition and under 808 nm light illumination with different intensities as shown in Figure 5b. The inset in Figure 5b shows the SEM image of the tested individual GeSe nanosheet device, which clearly indicates the uneven electrical contacts formation of the nanosheet with the Au electrodes. Figure S4 shows the AFM image of the device, which reveals the thickness of the nanosheet as ~57 nm. The photocurrent-time (I-t) responses to the pulses incident NIR light (808 nm) with different light intensities during ON-OFF cycles are shown in Figure 5c. Time-resolved photocurrent at +4V bias under multiple ON-OFF cycles shows a sharp rise in photocurrent under laser irradiation and the recovery process of photocurrent has a quick process followed by a slow recovery process under laser-off transition. The recovery process in photocurrent has rapid descent initially followed by a very slow decay component, which never reaches to the original level in short time. Similar phenomena have been observed for other photodetectors including single ZnO nanowire Schottky barrier photodetector. For nanosheet SB photodetector, the readsorbed ions (i.e. $O_2^-$) and relaxed carriers in SB interface are believed to be the major reason for the rapid recovery process of the GeSe SB photodetector.[28] The appearance of slow decay tail in the Schottky photodetector device is likely due to photogenerated carriers trapping at the interface between the nanosheet-Au electrodes. Figure 5d shows such two-step recovery process of the nanosheet based device. Under global illumination, when the full MSM device is illuminated with NIR light source, the IV response of the device shows asymmetric, nonlinear and unsaturated photoresponse. The observation indicates that in this structure the charge conduction is dominated by Schottky contacts.[27,28] We have checked other possible current conduction like space charge limited conduction (SCLC)[29] which might affect the IV characteristics, but we have not obtained reasonable fitting parameters as shown in Figure S5. For two Ohmic contacts MSM geometry under light illumination, the photocurrent response shows generally linear curve property, where the photoresponse is mainly due to increase in the photogenerated carriers in semiconductor and the total applied voltage is mainly dropped across the semiconducting material.[30,31] The nonlinear curves under illumination (Figure 5b) indicate the photocurrent response is dominated by SBs formation with the nanosheets. In this Schottky device, both the forward and reverse biased current increase under global laser light illumination. This implies photoexcited electron-hole (e-h) pairs significantly increase the concentration of majority carriers of the semiconductor in MSM structure, where the SBs heights also modulate with NIR illumination.[32,33]

The slow decay of photocurrent can be fitted with the following equation:

$$I(t) = I_d + A_1(e^{\frac{t_0-t}{\tau_d}}) \quad (1)$$

Where $I_d$ is the dark current, $t_o$ is the initial time, $A_1$ is the amplitude of the photocurrent, and $\tau_d$ is the response time of the decay curve. I-t curve shows two different time constants, $\tau \sim 0.1$s for the relative rapid rise (from OFF state to ON state), which is independent of the intensity of light illumination. Rapid decay time ($\tau_{d1} \sim 0.1$s) followed with relatively slow decay time $\tau_{d2}$ of $\sim 3.5$s (Figure 5d) is obtained under laser intensity of 283 mW/cm$^2$ (from ON state to OFF state). Such a slow photoresponce decay time observed in GeSe nanosheet may be due to presence of the deep trap states associated with various possible defects in the system.

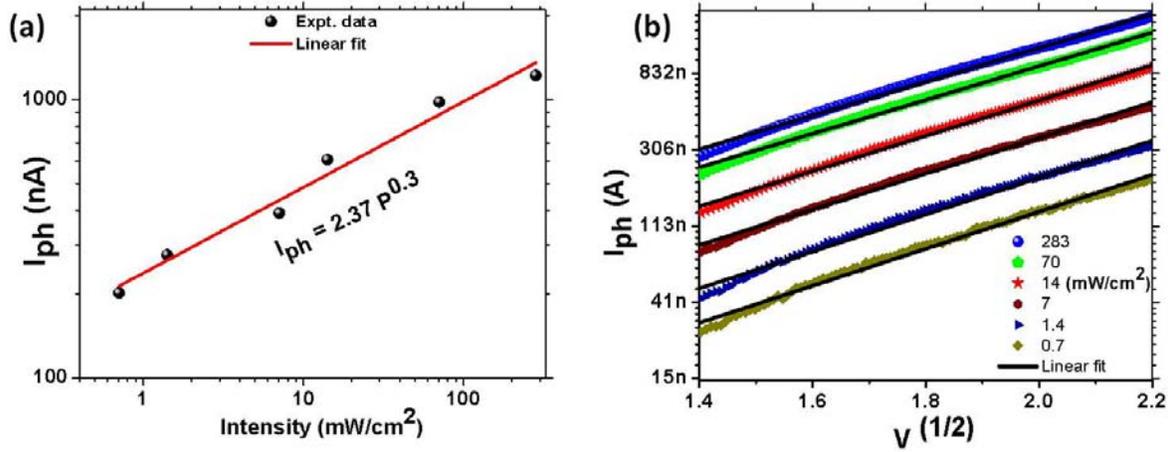

Figure 6. (a) Photocurrent as a function of light intensity under 808 nm and corresponding fitting curve using the power law. Both photocurrent and light intensity are in the log scale. (b) The plot of $I_{ph}$ (in log scale) with $V^{1/2}$ with different illuminated light intensities, and its fitted line (solid line).

Useful photodetectors often exhibit linear photoresponse with incident photon intensities. The power dependency of the photodetector has been investigated in Figure 6(a), which shows the photocurrent versus incident photon intensities under 808 nm-light illuminations as well as the fitting function using the power law. We used a simple power law: photocurrent, $I_{ph} = (A \times P^c)$ to fit the data, where A is proportionality constant, P is the light intensity irradiated on the device and c is the empirical coefficient and photocurrent $I_{ph} = (I-I_d)$ and I is the current of the device under illumination with the light source and $I_d$ is the dark current. Based on the experimental values, the fitting power law dependency to the experimental data gives A = 2.37 and c = 0.3. Fractional power dependence is believed to be related to carrier traps in the nanosheet, which are distributed within the energy gap.[34] The low value of exponent c of the photocurrent dependence on light intensity may be due to the fact that the trap states become recombination centers under illumination, leading to the weak light intensity dependence of photocurrent.[35] Here we adopt a similar strategy in analysis as Cheng et al.,[27] where the approximate photocurrent equation of ZnO nanowire based Schottky photodiode can be described as: $\ln(I_{ph}) \propto V^{1/2}$. Figure 6(b) shows the plot of $I_{ph}$ with $V^{1/2}$ in log scale with different intensities of 808 nm light irradiation and all curves are fitted with linear equation. These results indicate that our photocurrent responses of the devices are consistent with Schottky contacts dominated photoresponse of MSM structure.

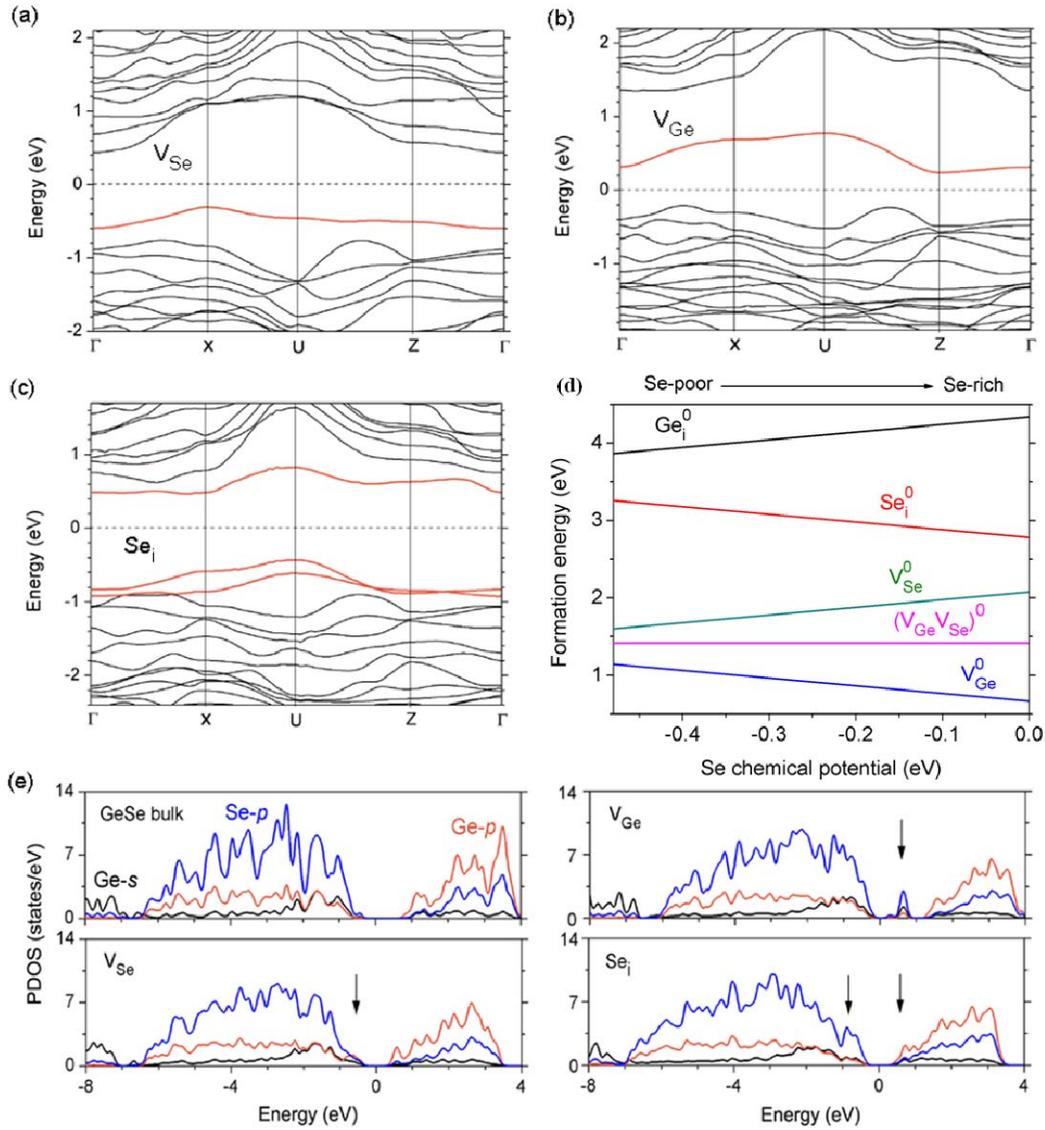

Figure 7. In-gap defective states (red lines) associated with three types of defects in GeSe : (a) $V_{Se}$, (b) $V_{Ge}$, (c) $Se_i$; (d) Formation energy of the neutral defects of GeSe as a function of the Se chemical potential, where the potential of Se rich condition corresponding to the energy of a single Se atom in $Se_6$ molecular crystal is set to zero. (e) PDOS for perfect bulk GeSe and the defective structures. The arrows denote the position of the defective states in the band gap.

Upon chemical tuning of the composition or physical treatment like light and heat during fabrication and application, intrinsic defects like atomic vacancies or interstitial atoms are created and ubiquitous in the GeSe matrix. The stability of the defects and the accompanying hole or electron states associated with the defects account for the different mobility and lifetime of the photo-induced carriers under different shining lights. Previous study of the photoconductivity of amorphous GeSe indicates a correlation between the magnitude of the photocurrent and the density of photoinduced defects.[36] In this study, first-principles calculations are employed to study the defects states associated with various defects in the sample. The calculations were performed using the Vienna *ab initio* simulation package (VASP).[37] Spin-restricted calculations using the projector-augmented wave potentials were conducted with the hybrid functional (HSE06),[38] in which part of the semilocal exchange-correlation functional with the generalized gradient approximation (GGA) is substituted by the Hartree-Fock exchange. We used a plane-wave basis set with a cutoff of 300 eV and integrations over the Brillouin zone were performed using a $3 \times 3 \times 2$ mesh of special *k* points. Defect calculations have been carried out for defective structures by creating defects on a $2 \times 2 \times 1$ supercell of GeSe containing 32 atoms. The structures were relaxed by adopting GGA functional until the force was less than 0.02 eV/A$^{-1}$. Formation energy ($E_f$)[39] of various defects was calculated by

$$E_f = E_{Defect} - E_{Perfect} - \sum_i n_i \mu_i \qquad (2)$$

Where $E_{Defect}$ and $E_{Perfect}$ are the total energy derived from a defective structure and equivalently perfect bulk, respectively. The defect is formed by adding or removing of $n_i$ atoms for the *i* type of atom with a chemical potential of $\mu_i$. In the extreme Se rich condition, the chemical potential of Se is equal to that in the Se$_6$ molecular crystal ($R\overline{3}$ phase). In the extreme Se poor limit, the chemical potential of Se is bounded by the formation of Ge in the film. We consider four types of defects: Se vacancy (V$_{Se}$), Ge vacancy (V$_{Ge}$), interstitial Se(Se$_i$), and a defect complex (V$_{Ge}$V$_{Se}$). We have also studied the local density of states calculation of pristine GeSe and O-adsorbed GeSe, which is described in Figure S6. The band structures of the supercell containing one defect center are shown in Figure 7(a-c). The formation energy of describing the energy needed for creating each defect is plotted in Figure 7(d). The partial density of states (PDOS) is shown in Figure 7(e). For perfect GeSe, our calculated value of an indirect band gap of 1.24 eV is in good agreement with experimental study of 1.07-1.29 eV.[40] The valence band is mainly composed of hybridized states formed between Ge-4*s*, -4*p*, and Se-4*p* orbital. The conduction band has larger component of 4*p* orbital of the cation than the 4*p* orbital of the anion. For the V$_{Se}$ defect, one occupied defective state near the valence maximum is formed. For the V$_{Ge}$ defect, it has the smallest formation energy. This is consistent with the fact that GeSe often shows nonstoichiometrical structures with an excess of Se. There is one deep unoccupied defective state. In the case of $Se_i$ defect, the interstitial Se is bonded to three Ge atoms distributed in the two adjacent GeSe planes, which is different from the $Se_i$ defect in $GeSe_2$ case where the interstitial Se atom is bonded to the two-coordinated Se atom forming a double bond with the Se 4$p$ lone-pair electrons. Two shallow occupied states are formed, whereas one shallow unoccupied state is formed below the conduction band. The PDOS plot indicates that the defective bonding and antibonding states are formed by hybridization between Ge-4$s$ and Se-4$p$ states. The presence of these shallow and deep defective states in the gap, facilitating the trapping and releasing of photo-excited carriers, can account for the slow photoresponse observed in Schottky photodiode under 808 nm light illumination and for the weak light intensity dependence of photocurrent of the Schottky photodetector.[21]

To analyze the NIR photodetecting performance, we have calculated the spectral responsivity ($R_\lambda$)[41] and the external quantum efficiency (EQE) or gain for the Schottky photodetectors. The large values of $R_\lambda$ and EQE refer high sensitivity of the photodetectors. Spectral responsivity is denoted as $R_\lambda = I_{ph}/(S \times P_\lambda)$; where S is the effective illuminated area, $P_\lambda$ is the light intensity. The dimensions of nanosheet, which are used for extracting the parameters of the photodetector, are channel width ~ 20 μm; height ~ 57 nm; and length ~ 10 μm. The maximum responsivity of the devices at fixed 4V external bias was estimated to be ~ 3.5 A W$^{-1}$ under 808 nm-light illumination (for a fixed laser intensity of 283 ± 0.1 mW/cm$^2$). External quantum efficiency (EQE),[42,43] an important parameter for photodetectors, is defined as EQE = (hc/eλ) × $R_\lambda$, where h is the Planck's constant, c is the speed of light, e is the electronic charge and λ is the excitation wavelength. The EQE of the single GeSe nanosheet device is estimated to be ~ 5.3 ×10$^2$ % at 4V fixed bias. The UV photodetector of ZnO nanowire Schottky barrier with high sensitivity shows photocurrent gain of 8.5 × 10$^3$ at 5V fixed bias.[28] Thus GeSe nanosheet based SB NIR photodetector shows poorer photocurrent gain with respect to ZnO nanowire based SB photodetector[28] and comparable to visible-blind deep-ultraviolet Schottky photodetector based on individual $Zn_2GeO_4$ nanowire device.[44]

The performance parameters of the GeSe nanosheet photodetector is compared in Table 1 to other 2D nanosheet photodetectors. GeSe nanosheet photodetector shows a comparable responsivity with several other reported 2D nanosheet devices.[45-47] 2D materials can avoid several limitations of 1D and 0D materials since they are established with design devices in semiconductor industry. These results indicate that GeSe nanosheet can be used as good sensitive nanoscale NIR photodetectors.

TABLE 1. Comparison with the Reported Parameters for 2D-nanostructure Photodetectors

| Photodetectors | Responsivity ($R_\lambda$) [A W$^{-1}$] | Quantum Efficiency (QE) (%) | Response time | Reference |
|---|---|---|---|---|
| few-layer GaSe | 2.8 | 1367 | 20 ms | 45 |
| single layer MoS$_2$ | 7.5×10$^{-3}$ | - | 50 ms | 46 |
| GaS nanosheet | 4.2 | 2050 | <30 ms | 47 |
| few-layer GeSe | 3.5 | 537 | 100 ms | This work |

**Conclusions**

In summary, high-quality single-crystalline GeSe nanosheets have been synthesized using chemical vapor deposition technique. Different shaped GeSe NSs were obtained during synthesis. Multilayer GeSe NSs films were deposited onto Si/SiO$_2$ substrates using the scotch tape-based mechanical exfoliation technique. Metal Au electrodes were deposited on GeSe NS to form two probe based device which shows SBs formation at contacts in Au-GeSe-Au structure. Two terminal based Schottky photodetector of GeSe nanosheet exhibits high photocurrent gain at 4V external bias. A maximum photoresponce was achieved of the devices with 808 nm-light illuminations. The maximum spectrum responsivity of ~ 3.5 A.W$^{-1}$ was obtained at 4V bias. The photoresponse and good responsivity suggest its potential applications as nanoscale photodetectors and photoelectric switches.

## Acknowledgements

This work is based on research supported by the NUS MOE FRC Grant No. R-144-000-281-112.

**Supporting Information Available:** XPS spectrum of GeSe nanosheets with the survey of full spectrum and high resolution O1s, Ge3d and Se3d core-level peaks are shown in Figure S1. Figure S2 provides I-V characteristics and I-t response of GeSe bulk flake device under 808 nm laser excitation. Incident photon to converted electron (IPCE) spectra of GeSe bulk flake device at zero volt external bias and the reflection spectrum of the GeSe nanosheets sample are shown in Figure S3. Trapping-mode AFM image and 3D AFM topography of the GeSe nanosheet device are shown in Figure S4, where the line profile shows the thickness of the nanosheet. In Figure S5, dark I-V curve with SCLC fitted curve is shown. Atomic model of interstitial 'O' species in GeSe and the local density of states for O-absorbed GeSe are shown in Figure S6. This material is available free of charge via the Internet at http://pubs.acs.org.

# Supporting Information

# NIR Schottky Photodetectors Based on Individual Single-Crystalline GeSe Nanosheet

Bablu Mukherjee, Yongqing Cai, Hui Ru Tan, Yuan Ping Feng, Eng Soon Tok*, and Chorng Haur Sow*

Department of Physics, 2 Science Drive 3, National University of Singapore (NUS), Singapore-117542

*E-mail: physowch@nus.edu.sg, phytokes@nus.edu.sg

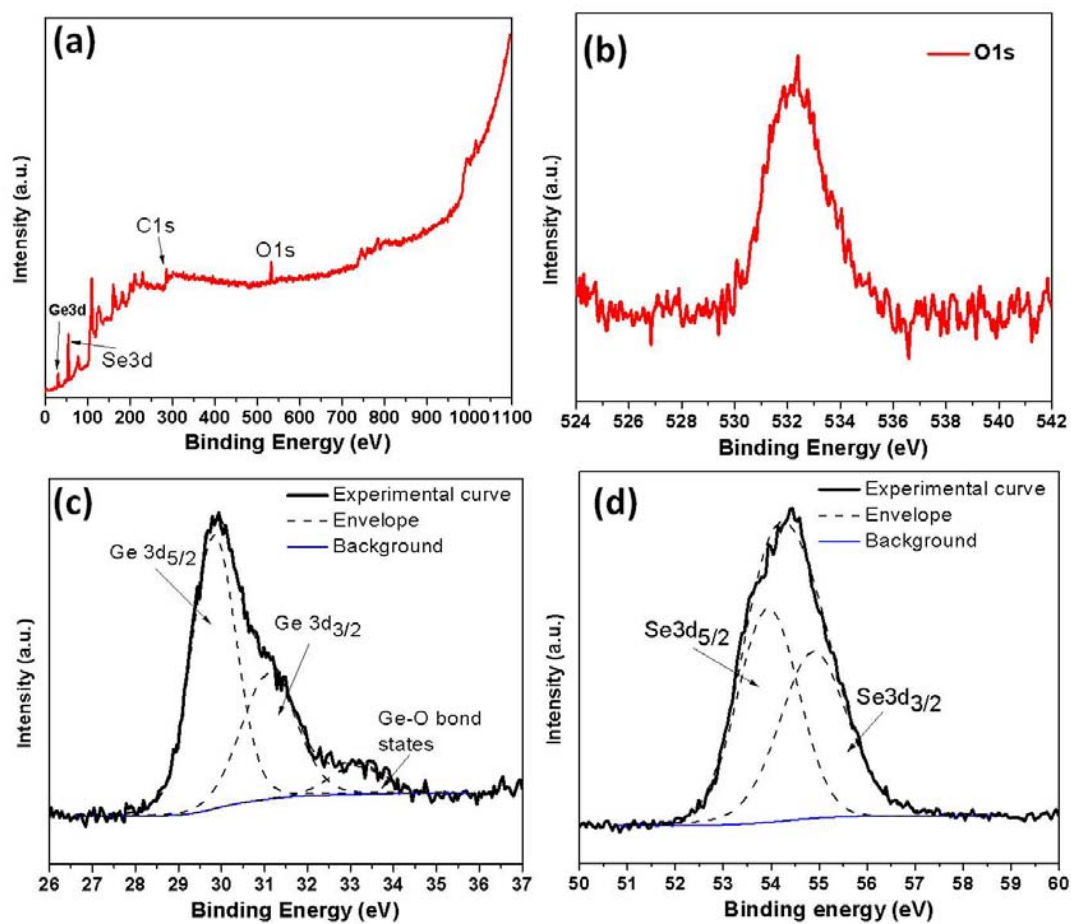

Figure S1. XPS spectrum of GeSe nanosheets. (a) Survey of full XPS spectrum. (b), (c) and (d) high resolution spectrum of O 1*s*, Ge 3*d* and Se 3*d*, respectively.

X-ray photoelectron spectroscopy (XPS) measurement was performed with a large area of as synthesized micron sized GeSe nanosheets. Full XPS spectrum of GeSe nanostructures is shown in Figure S1(a). The energy distributions of O 1$s$, Ge 3$d$ and Se 3$d$ photoelectrons are shown in Figure S1 (b-d). The peaks at ~ 29.8, 31.2, and 33.3 eV represent emission from Ge 3$d_{5/2}$, 3$d_{3/2}$, and Ge-O bond states levels from the Ge atoms while the peaks at ~ 53.9 and 54.9 eV come from Se 3$d_{5/2}$ and 3$d_{3/2}$ levels from the Se atoms (Figure S1.c,d). Thus O-related Ge-O bond states are observed on the surface of the GeSe nanostructures however we have not observed any Se-O bond states in the high resolution Se 3$d$ spectrum. Interestingly, binding energy associated with homopolar bonds Ge–Ge (at B. E. of 29.1 eV) and Se–Se (at B. E. of 55.4 eV) are not detected in our nanostructures.

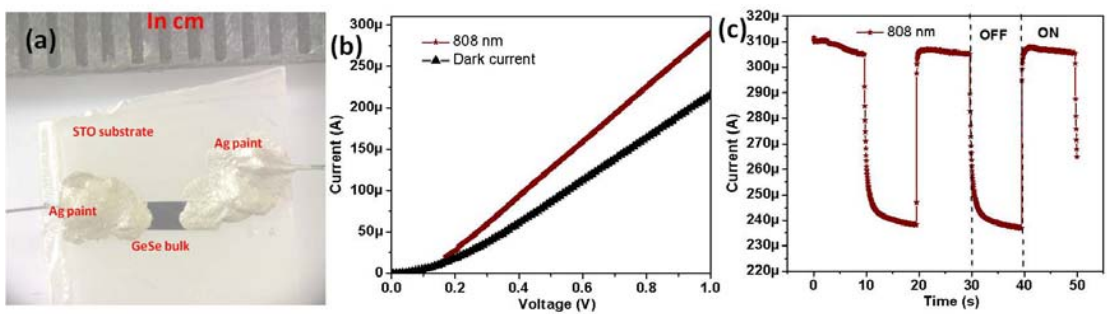

Figure S2. (a) GeSe bulk flake device made on STO substrate and the electrical contacts between GeSe bulk flake and electrical wires were made using silver paint. (b) The photoresponse of GeSe bulk flake device under laser excitation of 808 nm (fixed laser intensity of ~ 80 mW/cm$^2$). (c) Photocurrent-time (I-t) response of the bulk GeSe flake at fixed laser intensity of ~80 mW/cm$^2$.

The photo-response properties of bulk GeSe flake based photodetector were examined at 0V external bias. The photoresponse of the GeSe bulk flake device (Figure S2a) shows good photocurrent produce under illumination of NIR light as depicted in Figure S2b. The photocurrent-time (I-t) curves under 808 nm light illumination with OFF and ON states of light is shown in Figure S2c. I-t curves demonstrates higher photocurrent obtained under 808 nm illumination.

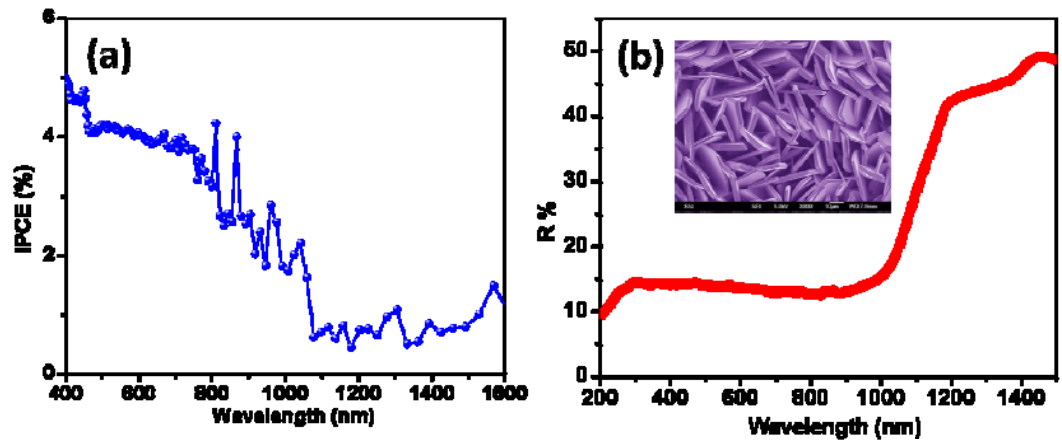

Figure S3. (a) Measured IPCE spectra of GeSe bulk flake device at the incident wavelength range from 400 to 1600 nm at a fixed zero bias. (b) Reflection spectrum of micron sized GeSe nanosheets. Inset shows false color SEM image of the sample.

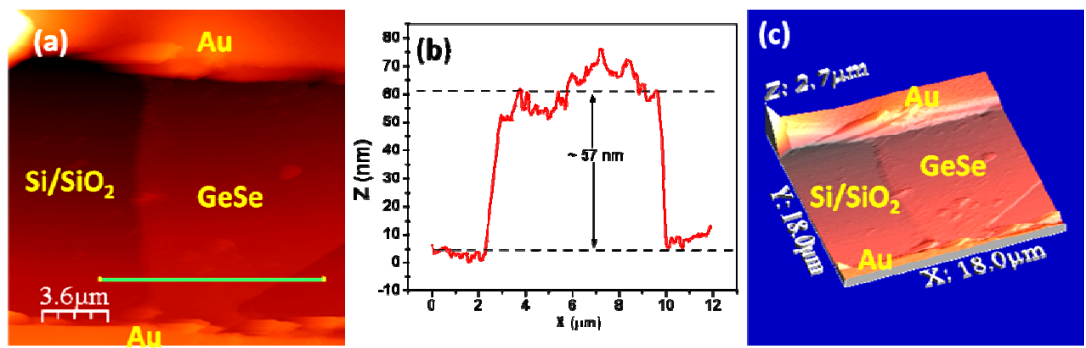

Figure S4. (a) Tapping-mode AFM image of a GeSe nanosheet bridging deposited Au electrodes. (b) The line profile taken along the green line of figure (a) shows the thickness of the GeSe nanosheet is ~ 57 nm. (c) 3D AFM topography of the GeSe nanosheet device.

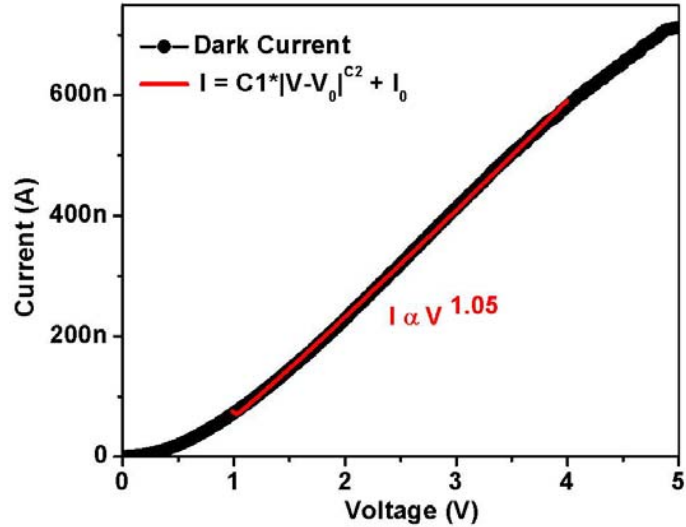

Figure S5. Dark I-V curve and solid red line represent the fitted curve with equation: I α $V^{C2}$, where C2 is a constant. The value of C2 is ~1.05, which is obtained from the fitted curve.

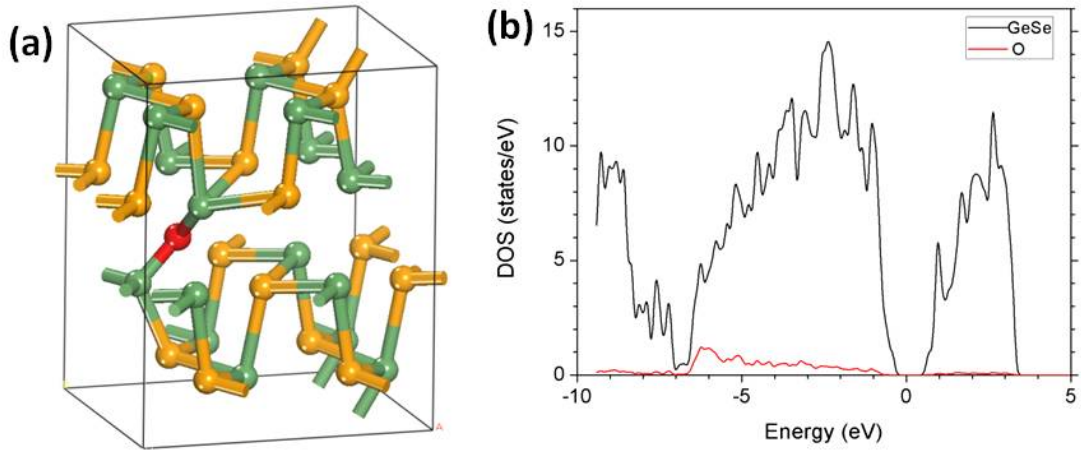

Figure S6 (a) Atomic model of interstitial 'O' species in GeSe. The green, yellow, and red balls represent Ge, Se, and O atoms, respectively. (b) Local density of states for O-adsorbed GeSe calculated by hybrid functional.

To estimate the effect of 'O' contamination in the sample, we take another hybrid functional calculation (Figure S6 a,b). It can be seen that there is no defective states related O in the gap of GeSe host. This suggests that the trapping of carriers due to 'O' contamination is unlikely to occur due to the absence of localized states in the gap. However, the presence of 'O' at the interface of the GeSe/Au contact plays important role in affecting the dipole at the interface, which will strongly affect the SBH and thus influencing the carriers' injection and diffusing across the junction.